\documentclass[preprint,showpacs,preprintnumbers,eqsecnum,amsmath,amssymb]{revtex4}
\usepackage{graphics,epsfig}
\usepackage{graphicx}
\usepackage{dcolumn}
\usepackage{bm}

\newcommand{\be}{\begin{equation}}
\newcommand{\ee}{\end{equation}}
\newcommand{\ba}{\begin{eqnarray}}
\newcommand{\ea}{\end{eqnarray}}

\begin{document}

\title{{\bf Late-time tail of a coupled scalar field
in the background of a black hole with a global monopole }}
\author{Songbai Chen}\email{csb3752@hotmail.com}
\author{Jiliang Jing} \email{jljing@hunnu.edu.cn}
\affiliation{ Institute of Physics and  Department of Physics, \\
Hunan Normal University,  Changsha, Hunan 410081, China\\ }

\begin{abstract}

Using the technique of spectral decomposition, we investigated the
late-time tails of massless and massive coupled scalar fields in
the background of a black hole with a global monopole. We found
that due to existence of the coupling between the scalar and
gravitational fields, the massless scalar field decay faster at
timelike infinity $i_+$, and so does the massive one in the
intermediate late time. But the asymptotically late-time tail for
the massive scalar field is not affected and its decay rate still
is $t^{-5/6}$.

\end{abstract}

 \pacs{ 04.30.-w, 04.62.+v, 97.60.Lf}
 \maketitle

\section{Introduction}
It is well known that the dynamical evolution of field
perturbations on a black hole consists roughly of three stages
\cite{Frolov98}. The first one is an initial wave burst coming
directly from the source of perturbation and is dependent on the
initial form of original wave field. The second one involves the
damped oscillations named quasinormal modes, which are dependent
only on the structure of the background spacetime and irrelevant
of the initial perturbation. The last one is the power-law tail
behavior of the field caused by backscattering of the
gravitational field at late time.

The study of the late-time behaviors of the fields outside black
holes has attracted a lot of attention for a long time
\cite{Wheeler1}-\cite{Jing3}. Price \cite{Price1} found that in
the Schwarzschild black hole spacetime the late-time behaviors of
the massless scalar, gravitational and electromagnetic
perturbations fields for a fixed $r$ are dominated by the factor
$t^{-(2l+3)}$.  In other static spacetimes, it is  found that the
late-time tails of these massless perturbations also possess the
same decay rate \cite{Hod2}-\cite{Cardoso}. Comparing with the
massless case, the late-time tail of the massive scalar field has
quite different characteristic. For example, it is dominated by
the oscillatory inverse power-law form $t^{-(l+3/2)}\sin{\mu t}$
in the intermediate late time \cite{Koyama}-\cite{Hod4}. The
similar properties of the late-time behaviors of these
perturbations were found in the rotational background metrics
\cite{Hod3}-\cite{Jing2}.

However, it is possible that the late-time behaviors of the
perturbations have some distinct properties in some special case.
For example, Jing \cite{Jing1} \cite{Jing3} recently considered the
evolution of the massive Dirac field in the Schwarzschild black hole
and of the massive charged Dirac field in the Reissner-Nordstr\"{o}m
black hole, and found that they are dominated by decaying tails
without any oscillation, which is quite different from that of the
scalar fields. Yu \cite{Hong} studied the late-time tail of the
scalar field in the black hole spacetime with a global monopole
\cite{Barriola1} and found that due to present of a solid angle
deficit, the massive scalar fields decay faster in the intermediate
times.

It is well known that the coupling between the scalar and
gravitational fields plays some important roles in the black hole
physics. In Ref.\cite{csb1}\cite{csb2}, we found the couple factor
affects the asymptotic quasinormal frequencies in the dilaton
black hole spacetimes. Then it is interesting to ask the question
whether the couple factor influence on the late-time tail of the
scalar field. Obviously, in the Schwarzschild,
Reissner-Nordstr\"{o}m, Kerr, and Kerr-Newman black holes, the
couple constant does not affect these late-time behaviors because
in which the Ricci scalar curvatures are zero and then the couple
terms vanish in the effective potentials. For the dilaton black
hole spacetimes, although the Ricci scalar curvatures are not
equal to zero, the couple factor still does not affect the
late-time behaviors because the couple term appears in the higher
order term $O(1/r^3)$. However, Brady \cite{Brady} found that in
the Schwarzschild de Sitter spacetime the late-time behaviors of
the coupled scalar field depend on the couple constant $\xi$.

Our purpose in this paper is to extend the initial works of Yu
\cite{Hong} and Brady \cite{Brady}. We consider the late-time
tails of the massless and massive coupled scalar fields in the
background of a black hole with a global monopole \cite{Barriola1}
and obtain some new results.

The plan of this paper is as follows. In Sec.II we present the
evolution equation of the coupled scalar field in the
 background of a black hole with a global monopole and introduce the
black hole Green's function using the spectral decomposition
method \cite{Morse}. In Sec.III and Sec.IV, we study the late-time
behavior of the massless and massive coupled scalar field in the
 spacetime with a global monopole, respectively. In
Sec.V we make a summary and some discussions.

\section{description of the system and black hole Green's function }
The metric of the  background of a black hole with a global
monopole is \cite{Barriola1}
\begin{eqnarray}
ds^2=-(1-8\pi G\eta^2_0-\frac{2Gm}{\bar{R}})d\tau^2+(1-8\pi
G\eta^2_0-\frac{2Gm}{\bar{R}})^{-1}d\bar{R}^2+\bar{R}^2(d\theta^2+
\sin^2{\theta}d\varphi^2)\label{m1},
\end{eqnarray}
where $m$ is the mass parameter of the black hole and $\eta_0$ is
the symmetry breaking scale when the monopole is formed.
Introducing the coordinate transformations
\begin{eqnarray}
t=(1-8\pi G\eta^2_0)^{\frac{1}{2}}\tau, \ \ \ r\rightarrow (1-8\pi
G\eta^2_0)^{-\frac{1}{2}}\bar{R},
\end{eqnarray}
and new parameters
\begin{eqnarray}
M=(1-8\pi G\eta^2_0)^{-\frac{3}{2}}m, \ \ \ b=(1-8\pi G\eta^2_0),
\end{eqnarray}
then  metric (\ref{m1}) can be rewritten as
\begin{eqnarray}
ds^2=-(1-\frac{2GM}{r})dt^2+(1
-\frac{2GM}{r})^{-1}dr^2+br^2(d\theta^2+\sin^2{\theta}d\varphi^2)
\label{m2}.
\end{eqnarray}
It is a static and spherically symmetric metric with an additional
solid angle deficit ($\Delta=4\pi b=32\pi G\eta^2_0$).

The Klein-Gordon equation for a coupled scalar field with mass
$\mu$  is
\begin{eqnarray}
 \frac{1}{\sqrt{-g}}\partial_\mu(\sqrt{-g}g^{\mu\nu}\partial_\nu)\phi
 -(\mu^2+\xi R)\phi=0.\label{eq1}
 \end{eqnarray}
where $\phi$ is the scalar field and $R$ is the Ricci scalar
curvature. The coupling between the scalar field and the
gravitational field is represented by the term $\xi R\phi$, where
 $\xi$ is a coupling constant.

Decomposing the field into
$\phi=\sum_{l,m}\frac{\psi(t,r)}{r}Y(\theta,\varphi)$ and
introducing the ``tortoise coordinate" change
  \begin{eqnarray}
  r_*=r+2M \ln{\frac{r-2M}{2M}},\label{x2}
\end{eqnarray}
we obtain a wave equation
 \begin{eqnarray}
\psi_{,tt}-\psi_{,r_*r_*}+V\psi=0\label{e1},
\end{eqnarray}
with the effective potential
\begin{eqnarray}
V=\left(1-\frac{2M}{r}\right)
\left[\frac{l(l+1)}{br^2}+\frac{2M}{r^3}+\mu^2 +\xi R\right],
\end{eqnarray}
where the Ricci scalar curvature is given by
\begin{eqnarray}
R=\frac{2(1-b)}{br^2}.
\end{eqnarray}
The time evolution of a wave field $\Psi(r_*,t) $ described by
Eq.(\ref{e1}) follows form \cite{Morse}
\begin{eqnarray}
\Psi(r_*,t)=\int{[G(r_*,r'_*;t)\partial_t\Psi(r'_*,0)+
\partial_t G(r_*,r'_*;t)\Psi(r'_*,0)]dr'_*},
\end{eqnarray}
where the retarded Green's function $G(r_*,r'_*;t)$ is defined by
\begin{eqnarray}
\left[\frac{\partial^2}{\partial t^2}-\frac{\partial^2}{\partial
r^2_*}+V(r)\right]G(r_*,r'_*;t)=\delta(t) \delta(r_*-r'_*).
\end{eqnarray}
The causality condition gives us the initial condition
$G(r_*,r'_*;t)=0$ for $t<0$. In order to get $G(r_*,r'_*;t)$, we
use the Fourier transform
\begin{eqnarray}
\tilde{G}(r_*,r'_*;\omega)=\int^{\infty}_{0^-}{G(r_*,r'_*;t)e^{i\omega
t}dt}.
\end{eqnarray}
The Fourier transform is analytic in the upper half
$\omega$-plane, and corresponding inversion formula is given by
\begin{eqnarray}
G(r_*,r'_*;t)=\frac{1}{2\pi}\int^{\infty+ic}_{-\infty+ic}{\tilde{G}(r_*,r'_*;\omega)e^{-i\omega
t}d\omega},\label{G1}
\end{eqnarray}
where $c$ is some positive constant. We define two auxiliary
functions $\tilde{\Psi}_1(r_*,\omega)$ and
$\tilde{\Psi}_2(r_*,\omega)$ which are linearly independent
solutions to the homogeneous equation
\begin{eqnarray}
\left[\frac{d^2}{dr^2_*}+\omega^2-V[r(x)]\right]\tilde{\Psi}_i(r_*,\omega)=0,\
\ \ i=1,2.\label{e2}
\end{eqnarray}
Let the Wronskian be
\begin{eqnarray}
W(\omega)=W(\tilde{\Psi}_1,\tilde{\Psi}_2)=\tilde{\Psi}_1\tilde{\Psi}_{2,r_*}
-\tilde{\Psi}_2\tilde{\Psi}_{1,r_*},
\end{eqnarray}
and using the solutions $\tilde{\Psi}_1$ and $\tilde{\Psi}_2$, the
black hole Green's function can be constructed as
\begin{eqnarray}
\tilde{G}(r_*,r'_*;\omega)=-\frac{1}{W(\omega)}\{
\begin{array}{c}
 \tilde{\Psi}_1(r_*,\omega)\tilde{\Psi}_2(r'_*,\omega),\ \ \ r_*<r'_*;\\
\tilde{\Psi}_1(r'_*,\omega)\tilde{\Psi}_2(r_*,\omega),\ \ \
r_*>r'_*.
\end{array}
\end{eqnarray}
To calculate $G(r_*,r'_*; t)$ using Eq.(\ref{G1}), one must close
the contour of integration into the lower half of the complex
frequency plane. It is by now well known that there exists a
branch cut in $\tilde{\Psi}_2$ placed along the negative imaginary
$\omega$-axis and the contribution to late-time tail comes from
the integral of $\tilde{G}(r_*,r'_*; \omega)$ around this branch
cut which is denoted by $G^C(r_*,r'_*; t)$. Thus, in the study of
late-time evolution of an external field, we just consider
$G^C(r_*,r'_*; t)$.

\section{late-time behavior of the massless coupled scalar field}

Let us now discuss the late-time behaviors of the massless coupled
scalar fields. It is well known that the late-time behavior of
massless perturbation fields is determined by the backscattering
from asymptotically far regions and the leading contribution to
the Green's function comes from the low-frequency parts. Thus, we
can study the asymptotic late-time behavior of the field by
adopting the low-frequency approximation. Neglecting terms of
order $O( (\frac{M}{r})^2 )$ and higher term, we can expend the
wave equation (\ref{e2}) for the massless scalar field as a power
series in $M/r$
\begin{eqnarray}
\left[\frac{d^2}{dr^2}+\omega^2+\frac{4M\omega^2}{r}-\frac{l(l+1)+2\xi
(1-b)}{br^2}\right]\zeta(r,\omega)=0,
\end{eqnarray}
where $\zeta(r,\omega)=\sqrt{1-\frac{2M}{r}}\tilde{\Psi}$. Let us
now introduce a second auxiliary field $\Phi(z)$ defined by
\begin{eqnarray}
&\zeta(r,\omega)&=r^{\rho+\frac{1}{2}}e^{-\frac{z}{2}}\Phi(z),
\end{eqnarray}
where
\begin{eqnarray}
z=-2i\omega r, \ \  \ \rho&=\sqrt{\frac{l(l+1)+2\xi
(1-b)}{b}+\frac{1}{4}}.
\end{eqnarray}
Then the equation becomes
\begin{eqnarray}
z\frac{d^2\Phi}{dz^2}+(1+2\rho-z)\frac{d\Phi}{dz}-(\frac{1}{2}+\rho
-2iM\omega)\Phi=0.
\end{eqnarray}
The two basic solutions of this equation required in order to
build the black hole Green's function are
\begin{eqnarray}
\tilde{\Psi_1}&=&Ae^{i\omega r}
r^{\frac{1}{2}+\rho}M(\frac{1}{2}+\rho-2iM\omega,1+2\rho,-2i\omega r),\\
\tilde{\Psi_2}&=&Be^{i\omega
r}r^{\frac{1}{2}+\rho}U(\frac{1}{2}+\rho-2iM\omega,1+2\rho,-2i\omega
r),
\end{eqnarray}
where $A$ and $B$ are normalization constants. $M(a, b, z)$ and
$U(a, b, z)$ are the two standard solutions to the confluent
hypergeometric equation. Since $U(a, b, z)$ is a many-valued
function, there will be a cut in $\tilde{\Psi_2}$. According to
Eq.(\ref{G1}), one finds that the branch cut contribution to the
Green's function is
\begin{eqnarray}
G^C(r_*,r'_*;t)&=&
\frac{1}{2\pi}\int^{-i\infty}_{0}{\tilde{\Psi}_1(r'_*,\omega
)\left[\frac{\tilde{\Psi}_2(r_*,\omega e^{2\pi i})}{W(\omega
e^{2\pi i})}-\frac{\tilde{\Psi}_2(r_*,\omega)}{W(\omega)}\right]
e^{-i\omega t} d\omega}.
\end{eqnarray}
Using the relation as follows
\begin{eqnarray}
\tilde{\Psi}_1(r_*,\omega e^{2\pi i})&=&\tilde{\Psi}_1(r_*,\omega
),\nonumber
\\ \tilde{\Psi}_2(r_*,\omega e^{2\pi i})&=&\frac{B}{A}\frac{2\pi ie^{-\pi(2\rho+1)i}}
{\Gamma{(2\rho+1)}\Gamma{(\frac{1}{2}-\rho-2i\omega
M)}}\tilde{\Psi}_1(r_*,\omega)+e^{-2\pi(2\rho+1)i}\tilde{\Psi}_2(r_*,\omega)\label{wf1},
\end{eqnarray}
we obtain
\begin{eqnarray}
W(\omega)=\frac{AB\;\Gamma{(2\rho+1)}(-i)^{-(2\rho+2)}(2\omega)^{-2\rho}}
{\Gamma{(\frac{1}{2}+\rho-2i\omega M)}}\label{w1},
\end{eqnarray}
and
\begin{eqnarray}
W(\omega e^{2\pi i})=e^{-2\pi i(2\rho+1)}W(\omega)\label{w2}.
\end{eqnarray}
In the low frequency approximation, the Green's function can be
expressed as
\begin{eqnarray}
G^C(r_*,r'_*;t)&=\frac{2^{2\rho}M[\Gamma{(\rho+\frac{1}{2})}]^2(-i)^{-2\rho}}{\pi
A^2[\Gamma{(2\rho+1)}]^2}\nonumber\\
&&\times
\int^{-i\infty}_0\tilde{\Psi}_1(r_*,\omega)\tilde{\Psi}_1(r'_*,\omega)\omega^{2\rho+1}e^{-i\omega
t}d\omega.\label{Gcf}
\end{eqnarray}
Let us now consider the asymptotic behavior at timelike infinity
$i_+$. As we described before, the late-time behavior of the field
arises from the low-frequency contribution to the Green's
function. Thus, the effective contribution to the integral in Eq.
(\ref{Gcf}) should come from $|\omega|=O(1/t)$. In the condition
that $|\omega|=O(1/t)$, we have
\begin{eqnarray}
\tilde{\Psi}_1(r_*,\omega )\simeq Ar^{\rho+1/2}_*,\ \ \
\tilde{\Psi}_1(r'_*,\omega )\simeq Ar'^{\rho+1/2}_*.\label{wa1}
\end{eqnarray}
Substituting them into Eq.(\ref{Gcf}) and performing the
integration, we find the asymptotic behavior of the massless
scalar field at timelike infinity is described by
\begin{eqnarray}
G^C(r_*,r'_*;t)&=&\frac{2^{2\rho}M[\Gamma{(\rho+\frac{1}{2})}]^2(-1)^{2\rho+1}\Gamma{(2\rho+2)}}{\pi
[\Gamma{(2\rho+1)}]^2}(r'_*r_*)^{\rho+1/2} t^{-2\rho-2}.
\end{eqnarray}
Thus, for a coupled massless scalar field, the late-time tails are
dominated not only by the multiple moment $l$ and the symmetry
breaking scale $\eta_0$, but also by the coupling constant $\xi$.
The presence of $\xi$ makes the field decay more rapidly. It
implies that the interaction between the matter and gravitation
fields plays an important role in the late-time evolutions of the
matter fields in the background of a black hole with a global
monopole .

\section{late-time behavior of the massive coupled scalar field}

For the massive case, we also just evaluate the $G^c(r_*,r'_*;t)$
as in the massless one. But there exist some slightly differences
in the massive scalar field case. The integral of the Green's
function $G(r_*,r'_*;t)$ around the branch cut performs in the
interval $-\mu \leq \omega \leq \mu$ \cite{Koyama} rather than
along the total negative imaginary $\omega$-axis. Assuming that
both the observer and the initial data are situated far away from
the black hole so that $r\gg M$, we can expand the wave equation
(\ref{e2}) as a power series in $M/r$ and obtain [neglecting terms
of order $O((\frac{M}{r})^2)$ and higher]

\begin{eqnarray}
\left[\frac{d^2}{dr^2}+\omega^2-\mu^2+\frac{4M\omega^2-2M\mu^2}{r}-\frac{l(l+1)+2\xi
(1-b)}{br^2}\right]\zeta(r,\omega)=0\label{e3},
\end{eqnarray}
where $\zeta(r,\omega)=\sqrt{1-\frac{2M}{r}}\tilde{\Psi}$. The
equation can be rewritten as the confluent hypergeometric equation
\begin{eqnarray}
z\frac{d^2\Phi}{dz^2}+(1+2\rho-z)\frac{d\Phi}{dz}-(\frac{1}{2}+\rho
-\lambda)\Phi=0,
\end{eqnarray}
with
\begin{eqnarray}
&z&=2\sqrt{\mu^2-\omega^2r}=2\varpi r, \ \ \ \ \ \
\zeta(r,\omega)=z^{\rho+\frac{1}{2}}e^{-\frac{z}{2}}\Phi(z),\nonumber \\
&\lambda&=\frac{M\mu^2}{\varpi}-2M\varpi,\ \ \ \ \
\rho=\sqrt{\frac{l(l+1)+2\xi (1-b)}{b}+\frac{1}{4}},
\end{eqnarray}
and the two solutions required in order to build the Green's
function are given by (for $|\omega|\ll \mu $)
\begin{eqnarray}
\tilde{\Psi_1}&=&A'M_{\lambda,\rho}(2\varpi r)=A'e^{-\varpi
r}(2\varpi
r)^{\frac{1}{2}+\rho}M(\frac{1}{2}+\rho-\lambda,1+2\rho,2\varpi r),\\
\tilde{\Psi_2}&=&B'W_{\lambda,\rho}(2\varpi r)=B'e^{-\varpi
r}(2\varpi
r)^{\frac{1}{2}+\rho}U(\frac{1}{2}+\rho-\lambda,1+2\rho,2\varpi
r),
\end{eqnarray}
where $A'$ and $B'$ are normalization constants.

Making use of Eq.(\ref{G1}), one finds that the contribution
originated from the branch cut to the Green's function is given by
\begin{eqnarray}
G^C(r_*,r'_*;t)&=&
\frac{1}{2\pi}\int^{\mu}_{-\mu}{\left[\frac{\tilde{\Psi}_1(r'_*,\omega
e^{i \pi})\tilde{\Psi}_2(r_*,\omega e^{i\pi})}{W(\omega
e^{i\pi})}-\frac{\tilde{\Psi}_1(r'_*,\omega)\tilde{\Psi}_2(r_*,\omega)}{W(\omega)}\right]
e^{-i\omega t} d\omega}
\nonumber \\
&=&\frac{1}{2\pi}\int^{\mu}_{-\mu}{F(\varpi)e^{-i\omega t}
d\omega}.\label{G3}
\end{eqnarray}
Using the following relation
\begin{eqnarray}
W_{\lambda,\rho}(2\varpi r)&=&\frac{\Gamma (-2\rho)}{\Gamma
(\frac{1}{2}-\rho-\lambda)} M_{\lambda,\rho}(2\varpi
r)+\frac{\Gamma (2\rho)}{\Gamma (\frac{1}{2}+\rho-\lambda)}
M_{\lambda,-\rho}(2\varpi r),\\
M_{\lambda,\rho}(e^{i\pi}2\varpi
r)&=&e^{(\frac{1}{2}+\rho)i\pi}M_{-\lambda,\rho}(2\varpi r),
\end{eqnarray}
we find
\begin{eqnarray}
W(\varpi e^{i\pi})=-W(\varpi)=A'B'\frac{\Gamma (2\rho)}{\Gamma
(\frac{1}{2}+\rho-\lambda)}4\rho\varpi,
\end{eqnarray}
and
\begin{eqnarray}
&&F(\varpi)=\frac{1}{4\rho \varpi}[M_{\lambda,\rho}(2\varpi
r'_*)M_{\lambda,-\rho}(2\varpi r_*)-M_{-\lambda,\rho}(2\varpi
r'_*)M_{-\lambda,-\rho}(2\varpi r_*)]+\frac{1}{4\rho \varpi}\times
\nonumber\\ &&\frac{\Gamma (-2\rho)\Gamma
(\frac{1}{2}+\rho-\lambda)}{\Gamma (2\rho)\Gamma
(\frac{1}{2}-\rho-\lambda)}[M_{\lambda,\rho}(2\varpi
r'_*)M_{\lambda,\rho}(2\varpi
r_*)-e^{(2\rho+1)i\pi}M_{-\lambda,\rho}(2\varpi
r'_*)M_{-\lambda,\rho}(2\varpi r_*)].\label{F1}
\end{eqnarray}

\subsection{the intermediate late-time behavior}

Let us now focus on the intermediate late-time behavior of the
massive scalar field. That is the tail in the range $M\ll r\ll t\ll
M/(\mu M)^2$. In this time scale, it is very easy for us to verify
that the dominant contribution to the integral in the Green's
function $G^C(r_*,r'_*;t)$ arises from the frequency range
$\varpi=O(\sqrt{\mu/t})$, or equivalently $\lambda\ll 1$. From the
massive scalar field equation(\ref{e3}), we know that $\lambda$
originates from the $1/r$ term which describes from the effect of
backscattering off the spacetime curvature. Thus, the parameter
$\lambda\ll 1$ means that the backscattering off the curvature from
the asymptotically far regions is negligible.  Therefore, we have in
this case
\begin{eqnarray}
F(\varpi)\approx \frac{1+e^{(2\rho+1)i\pi}}{4\rho
\varpi}\frac{\Gamma (-2\rho)\Gamma (\frac{1}{2}+\rho)}{\Gamma
(2\rho)\Gamma (\frac{1}{2}-\rho)}M_{0,\rho}(2\varpi
r'_*)M_{0,\rho}(2\varpi r_*).\label{f1}
\end{eqnarray}
In order to calculate the intermediate late-time behavior of the
massive scalar field at a fixed radius (where $r'_*,r_*\ll t$), we
can make use of the limit $\varpi r\ll 1$ and the property
$M(a,b,z)\approx 1$ as $z$ approaches zero. Then we find that
Eq.(\ref{f1}) can be approximated as
\begin{eqnarray}
F(\varpi)\approx \frac{1+e^{(2\rho+1)i\pi}}{4\rho
2^{-2\rho-1}}\frac{\Gamma (-2\rho)\Gamma
(\frac{1}{2}+\rho)}{\Gamma (2\rho)\Gamma
(\frac{1}{2}-\rho)}(r'_*r_*)^{\frac{1}{2}+\rho}\varpi^{2\rho}\label{f2}.
\end{eqnarray}
Thus, in the limit $t\gg \mu^{-1}$, the Green's function
$G^C(r_*,r'_*;t)$ becomes
\begin{eqnarray}
G^C(r_*,r'_*;t)&=&\frac{(1+e^{(2\rho+1)i\pi})}{\pi \rho
2^{-3\rho-2}}\frac{\Gamma (-2\rho)\Gamma (\frac{1}{2}+\rho)\Gamma
(1+\rho)\mu^{\rho}}{\Gamma (2\rho)\Gamma (\frac{1}{2}-\rho)}\nonumber \\
&&\times(r'_*r_*)^{\frac{1}{2}+\rho}t^{-\rho-1}\cos{[\mu t-\frac{\pi
(\rho+1)}{2}]}.\label{G4}
\end{eqnarray}
We find the intermediate late-time behavior of a coupled massive
scalar field is dominated by an oscillatory inverse power-law tail
which decays slower than the massless case. As the massless case,
the power-law tail of a coupled massive field at a fixed radius in
the intermediate time  also depends on the multiple number of the
wave modes, the symmetry breaking scale $\eta_0$ and the coupling
between the scalar and gravitational fields. Furthermore, we find
that the coupled scalar field decays faster than the non-coupled
one \cite{Hong} in this black hole spacetime.

\subsection{the asymptotical late-time behavior}
We now discuss the asymptotical late-time tail of a coupled
massive scalar field. Since the asymptotic tail behavior is caused
by a resonance backscattering of spacetime curvature at very late
times $\mu t\gg 1/(\mu M)^2$, it is expected that the inverse
power-law decay is replaced by another pattern of decay. In this
case the backscattering from asymptotically far regions is
important because that the parameter $\lambda$ can not be
negligible. When $\lambda\gg 1$, we find
\begin{eqnarray}
M_{\pm\lambda,\pm\rho}(2\varpi r)\approx \Gamma(1\pm
2\rho)(2\varpi r)^{\frac{1}{2}}(\pm \lambda)^{\mp\rho}J_{\pm
2\rho}(\sqrt{\pm \alpha r}),
\end{eqnarray}
where $\alpha=8\lambda \varpi$. Consequently, Eq.(\ref{F1}) can be
expressed as
\begin{eqnarray}
F(\varpi)&=&
\frac{\Gamma(1+2\rho)^2\Gamma(-2\rho)\Gamma(\frac{1}{2}+\rho-\lambda)r'_*r_*}
{2\rho\Gamma(2\rho)\Gamma
(\frac{1}{2}-\rho-\lambda)}\lambda^{-2\rho}[J_{2\rho}(\sqrt{\pm
\alpha r'_*})J_{2\rho}(\sqrt{\pm \alpha r_*})\nonumber \\
&& +I_{2\rho}(\sqrt{\pm \alpha r'_*})I_{2\rho}(\sqrt{\pm \alpha
r_*})]+\frac{\Gamma(1+2\rho)\Gamma(1-2\rho)r'_*r_*}{2\rho
}\nonumber \\ && [J_{2\rho}(\sqrt{\pm \alpha
r'_*})J_{-2\rho}(\sqrt{\pm \alpha r_*})-I_{2\rho}(\sqrt{\pm \alpha
r'_*})I_{2\rho}(\sqrt{\pm \alpha r_*})],\label{F3}
\end{eqnarray}
where $I_{\pm2\rho}$ is the modified Bessel function. It is
obvious that the late-time tail arising from the second term has a
form $t^{-1}$. Now let us discuss the late-time behavior come from
the first term. To calculate conveniently, we define
\begin{eqnarray}
L&=& \frac{\Gamma(1+2\rho)^2\Gamma(-2\rho)r'_*r_*}
{2\rho\Gamma(2\rho)}[J_{2\rho}(\sqrt{\pm \alpha
r'_*})J_{2\rho}(\sqrt{\pm \alpha r_*})+I_{2\rho}(\sqrt{\pm \alpha
r'_*})I_{2\rho}(\sqrt{\pm \alpha r_*})].
\end{eqnarray}
The contribution of the first term in Eq.(\ref{F3}) to the Green's
function can be approximated as
\begin{eqnarray}
G^C_1(r_*,r'_*;t)=\frac{L}{2\pi}\int^{\mu}_{-\mu}\frac{1+(-1)^{2\rho}e^{-i2\pi
\lambda}}{1+(-1)^{2\rho}e^{i2\pi \lambda}}e^{i(2\pi\lambda-\omega
t)}d\omega.
\end{eqnarray}
Making use of the saddle-point integration \cite{Koyama}, we can
obtain the asymptotic tail arising from the first term is $\sim
t^{-5/6}$ and it dominates over the tail from the second term.
Therefore, we obtain $ G^C(r_*,r'_*;t)\sim t^{-5/6}$. Obviously,
the asymptotic late-time tail of a coupled massive scalar field is
an oscillatory tail with the decay rate of $t^{-5/6}$, which
agrees with that of the non-coupled massive scalar field and can
be regarded as a quite general feature for the late-time decay of
massive scalar field.

\section{summary and discussion}
In summary, we have studied analytically the late-time behaviors
of the coupled massless and massive scalar fields in the
background of a black hole with a global monopole . We find that
both the asymptotic late-time tail of the coupled massless scalar
field at timelike infinity and the intermediate late-time tail of
the coupled massive scalar field at a fixed radius depend not only
on the multiple moment $l$ and the symmetry breaking scale
$\eta_0$, but also on the couple constant $\xi$ between the scalar
and gravitational fields. We note that the larger the couple
constant $\xi$, the faster the decay of the scalar fields. When
$b=1$, our result returns that of in the general Schwarzschild
spacetime. The couple constant $\xi$ disappears in the late time
behavior because that the Ricci scalar curvature becomes zero.
When $b=0$, the background spacetime can not described by
Barriola-Vilenkin metric (\ref{m1}). Thus, the late-time behaviors
of external fields in this case need to be investigated in the
future.

Comparing our result with that of in the Schwarzschild de Sitter
spacetime \cite{Brady}, we find that in the Schwarzschild de Sitter
spacetime, although the late-time behavior of the massless field
depends on the value of the coupling constant $\xi$, the decay
factor is independent of $\xi$ when $\xi>\xi_c$ (where $\xi_c$ is a
critical value) \cite{Brady}. However, in the background of a black
hole with a global monopole, we find that the coupling constant
always plays role in the intermediate late-time decay of the coupled
scalar field.

\vspace*{0.1cm}
\begin{center}\textbf{Acknowledgements}\end{center}
This work was supported by the National Basic Research Program of
China under Grant No. 2003CB716300;  the National Natural Science
Foundation of China under Grant No. 10275024 and under Grant No.
10473004; the FANEDD under Grant No. 200317; the Hunan Provincial
Natural Science Foundation of China under Grant No. 04JJ3019; and
the Hunan Normal University  Natural Science Foundation Grant No.
22040639.


\newpage
\begin{thebibliography}{99}

\bibitem{Frolov98}V.P. Frolov and I.D. Novikov,
\textit{Black hole physics: basic concepts and new developments}
(Kluwer Academic publishers 1998)

\bibitem{Wheeler1}R. Ruffini and J.A. Wheeler, Phys. Today {\bf 24(1)}, 30 (1971)

\bibitem{Wheeler2}C.W. Misner, K.S. Thorne, and J.A.
Wheeler, \textit{gravitation} (Freeman, San Francisco, 1973).

\bibitem{Price1}R.H. Price, Phys. Rev. D {\bf 5}, 2419 (1972).

\bibitem{Hod2}S. Hod and T. Piran,
Phys. Rev. D {\bf 58}, 024017 (1998).

\bibitem{Cai}R.G. Cai and A.Z. Wang,
Gen. Rel. Grav. {\bf 31}, 1367-1382 (1999).

\bibitem{Barack2}L. Barack,
 Phys. Rev. D {\bf 61} 024026 (2000).

 \bibitem{Burko1}L.M. Burko and G. Khanna
Phys. Rev. D {\bf 67}, 081502 (2003).

\bibitem{Ching}E.S.C. Ching, P.T. Leung, W.M. Suen, and K. Young,
Phys. Rev. D {\bf 52}, 2118 (1995).

\bibitem{Brady3}P.R. Brady, S. Droz, and S.M. Morsink
 Phys. Rev. D {\bf 58} 084034 (1998).

\bibitem{Price2}C. Gundlach, R.H. Price, and J. Pullin,
Phys. Rev. D {\bf 49}, 883 (1994).

\bibitem{Burko}L.M. Burko and G. Khanna
Phys. Rev. D {\bf 70}, 044018 (2004)

\bibitem{Cardoso}V. Cardoso, S. Yoshida, O.J.C. Dias, and J.P.S. Lemos
Phys. Rev. D {\bf 68}, 061503(R) (2003)

\bibitem{Koyama}H. Koyama and A. Tomimatsu,
Phys. Rev. D {\bf 63}, 064032 (2001).

\bibitem{Koyama1}H. Koyama and A. Tomimatsu,
Phys. Rev. D {\bf 64}, 044014 (2001).

\bibitem{Moderski}R. Moderski and M. Rogatko,
Phys. Rev. D {\bf 64}, 044024 (2001).

\bibitem{Moderski1}R. Moderski and M. Rogatko, Phys. Rev. D {\bf 63}, 084014
(2001).

\bibitem{Moderski2}R. Moderski and M. Rogatko, Phys. Rev. D {\bf 72},  044027
(2005).


\bibitem{Leaver}E.W. Leaver,
Phys. Rev. D {\bf 34}, 384 (1986).
\bibitem{Hod4}S. Hod and T. Piran,
Phys. Rev. D {\bf 58}, 044018 (1998).

\bibitem{Hod3}S. Hod,
Phys. Rev. D {\bf 58}, 104022 (1998).

\bibitem{Barack5}L. Barack and A. Ori,
Phys. Rev. Lett. {\bf 82}, 4388 (1999).

\bibitem{Krivan}W. krivan,
Phys. Rev. D {\bf 60}, 101501(R) (1999).

\bibitem{Andersson}K. Glampedakis and N. Andersson
Phys. Rev. D {\bf 64}, 104021 (2001).

\bibitem{Jing2}Q. Y. Pan and J. L. Jing, gr-qc/0405129,
Chin. Phys. Lett. {\bf 21}, 1873 (2004).

\bibitem{Jing1}J. L. Jing, Phys.Rev. D{\bf 72}, 027501 (2005)



\bibitem{Jing3} J. L.  Jing,  Phys. Rev. D {\bf 70}, 065004 (2004).

\bibitem{Hong}H. Yu, Phys. Rev. D {\bf 65}, 087502 (2002).

\bibitem{Barriola1}M. Barriola and A. Vilenkin,
Phys. Rev. Lett. {\bf 63}, 341 (1984).

\bibitem{csb1} S. B. Chen and J.L. Jing, Class. and Quantum Grav. {\bf
22} 533 (2005)
\bibitem{csb2}S. B. Chen and J.L. Jing, Class. and Quantum Grav. {\bf
22} 2159 (2005)

\bibitem{Brady}P.R. Brady, C.M. Chambers, W.G. Laarakkers and
E. Poisson, Phys. Rev. D {\bf 60}, 064003 (1999).


\bibitem{Morse}P.M. Morse and H. Feshbach, \textit{Methods of
Theoretical Physics} (McGraw-Hill, New York, 1953).


\end{thebibliography}
\end{document}